\documentclass[12pt,a4paper]{article}
\usepackage{graphicx}
\usepackage[cp1251]{inputenc}
\usepackage{amssymb, amsfonts, amsmath}
\numberwithin{equation}{section}
\date{}
\usepackage{amscd}
\newtheorem{theor}{Theorem}

\newenvironment{proof}[1][Proof]
{\textbf{#1.} }{\ \rule{0.5 em}{0.5 em}}

\begin{document}

\title{Algebras of integrals of motion for the Hamilton-Jacobi and Klein-Gordon-Fock equations in spacetime with a four-parameter groups of motions in the presence of an external electromagnetic field}

 \author{V.V. Obukhov}

 \maketitle

 \quad

  Institute of Scietific Research and Development,

Tomsk State Pedagogical University (TSPU), 634061 Tomsk,  Russia

 \quad

  Laboratory for Theoretical Cosmology, International Centre of Gravity and Cosmos,

Tomsk State University of Control Systems and Radioelectronics (TUSUR), 634050 Tomsk

\quad


            and tensors, integrals of motion.

            \quad

\section{Introduction}

 There are currently two main methods of exact integration of the classical and quantum equations of a test particle motion (the Hamilton---Jacobi, Klein---Gordon---Fock, and Dirac---Fock equations) in flat and curved spacetime, including when the external electromagnetic field is present. Both methods are based on the algebra of integrals of motion, each of which is linear or quadratic in momenta.

Symmetry operators of commutative algebra form a complete set of integrals of motion. A necessary condition for this set’s existence is the presence of a complete set of Killing vector and tensor fields in the space. In this case, a complete separation of variables takes place.

 The complete separation of variables is based on the theory of Stackel spaces. Spaces in which the free Hamilton-Jacobi equation is integrated by the complete separation of variables are called Stackel spaces. The coefficients at the highest powers of the products of momenta included in the first integrals of motion are vector (if the integral is linear in momenta) or tensor (for integrals of motion quadratic in momenta) Killing fields. It is also true for the Hamilton-Jacobi and Klein-Gordon-Fock equations in an electromagnetic field. Thus, the symmetry of the classical and quantum equations of motion is associated with the symmetry of space. Methods of complete separation of variables for classical and quantum equations of motion of charged test particles in the presence of external fields are constructed within the framework of the theory of Stackel spaces. The methods have been successfully used to solve classification problems in general relativity since 1966. (see, for example, \cite{Sh1}-\cite{Bagrov1} and the links indicated therein).

 Otherwise, the non-commutative integration method is applied. The non-commutative integration method was developed for quantum one-particle equations of motion (Klein-Gordon-Fock, Dirac-Fock, etc.) In this case, the set is formed by the operators of the non-commutative group of space motions. Since these operators do not commute between themselves, they cannot be directly used to construct a basic solution.
 The original algebra is supplemented with linear differential operators acting in the space of functions depending on coordinates and parameters to get around this problem.
 Extended operators are selected in such a way that they mutually commute. Using them, one can find a basic solution.

 The problem of non-commutative integration attracted the attention of researchers much later, although the complete algebraic classification of spaces that admit groups of motions required for its statement was constructed by A.Z. Petrov back in the sixties of the last century (see \cite{Petrov}). The classification served as the basis for constructing a method for the exact integration of linear partial differential equations using non-commutative algebras of first-order integrals of motion (see \cite{ASh} -- \cite{ASh4}).

 The method made it possible to significantly expand the sets of external fields and   metrics of   spacetime,  in which  basic solutions of the quantum equations of motion for a charged test particle, can be found.  It became an incentive for the construction the set of admissible electromagnetic fields as a necessary stage in the application of a non-commutative integration procedure (electromagnetic
fields in which the Hamilton---Jacobi and (or) Klein---Gordon---Fock equations for a charged test particle admit algebra of integrals of motion linear in momenta are called admissible).

In particular, in \cite{Mag1}, \cite{Mag3} a classification of all potentials of admissible electromagnetic fields for  ((metrics of spacetime)) admitting a transitive action of four-parameter movement groups \quad $G_4$ \quad is constructed.  In \cite{OVV1},\cite{OVV1a} a similar classification was carried out for homogeneous spaces with a three-parameter groups of motions \quad $G_3$,\quad acting transitively on non null hyperspace of spacetime (see \cite{OVVA} too).

  In present work, the following is done.

  1. For the case of an $ n $ -dimensional Riemannian space with an arbitrary signature $ \quad V_n, \quad $ it is proved that, in an admissible electromagnetic field, the algebras of integrals of motion of the Hamilton-Jacobi equation and the symmetry operators of the Klein-Gordon-Fock equation for a charged particle coincide, and they coincide with the algebras of symmetry operators for the free Hamilton-Jacobi and Klein-Gordon-Fock equations.

  2. In the $n$-dimensional Riemannian space $\quad V_n,\quad $ conditions that must be satisfied by admissible electromagnetic fields are obtained, and the compatibility of the equations systems following from these conditions is proved for the case when the group of motions \quad $G_r$ \quad acts transitively on the subspace \quad $V_r \quad (r\leq n, \quad rank(G_r)=r $). \quad

  3.	The classification of admissible electromagnetic fields in spacetime with four-parameter groups of motions \quad $ G_4, $ \quad is completed for the case when \quad$ rank(G_4)<4 $.

  \quad

  Note that additional conditions on metrics and potentials (for example, the Einstein equations) are not used.

  \maketitle

  \section{Conditions for the existence of the algebra of integrals of motion.}
\subsection{Designations.}
 In the next two sections, we consider the $n$ - dimensional Riemannian space \quad $V_n$\quad (with an arbitrary signature), on the $r$-dimensional subspace \quad $V_r$\quad of which  the $R$-parameter groups of motions \quad $ G_R\quad (rank(G_R) = r \leq R )$ \quad of the space \quad $V_n$\quad acts transitively. This means that any two points of the subspace \quad $V_r$ \quad can be transformed into each other by some transformation of the group \quad $G_R$. \quad In this case, the subspace \quad $V_r$\quad is called the subspace of transitivity (or the hypersurface of transitivity if \quad $ r = n - 1 $).

 The following index notations are used. The coordinate indices of the canonical local coordinate system variables \quad $[u^i] $\quad of the space \quad $V_n$\quad will be denoted by small Latin letters:\quad $i,j,l = 1, \dots n$.\quad
By analogy with Stackel spaces, all the variables of the canonical coordinate system are divided into two sets.
Variables of the local coordinate system of the subspace \quad $V_r$,\quad on which
the group of motions \quad $G_R$ \quad acts transitively, are called ignored and are supplied with small
letters of the Greek alphabet: \quad $ \alpha, \beta, \gamma $\quad varying within \quad $1, \dots r.$\quad
Ignored variables enter the metric tensor only through specific functions.

The rest of the variables are called essential (or non-ignored). Coordinate indices of essential variables are denoted by the letters  \quad $ p, q =  1, \dots r$.  The essential  variables enter the metric tensor and  the components of the potential of an admissible electromagnet field
through arbitrary functions, each of which depends only on these variables.

If $R>r$, the group indices will be denoted by capital Latin letters: $\quad A, B, C, D = 1, \dots R. \quad $ The Independent operators \quad $X_A$ \quad of the group \quad $G_R$\quad are numbered, the same way as the local coordinates of the subspace \quad $V_r$, \quad in small Greek letters. The other operators of the group \quad $G_R$,\quad are linear combinations of the independent operators \quad $X_\alpha$\quad (with variable coefficients) and are numbered by the indices:\quad $a, b, c, d = (r+1), \dots, R $. \quad The coordinate indices of the essential variables of the local canonical coordinate system are denoted by the following letters: \quad $p, q =r+1,\dots ,n.$\quad  In the metric tensor and in the components of the admissible electromagnetic field potential, essential variables enter through arbitrary functions, each of which depends only on these variables. The repeated superscripts and subscripts are summed.
  The subject of our consideration is the conditions imposed on the external electromagnetic field, under which there are algebras of integrals of motion linear in momenta of the classical and quantum equations of motion for a charged scalar test particle.

  \subsection{ Derivation of the conditions for the existence of an admissible electromagnetic field.}

  For a charged test particle in an external electromagnetic field with potential \quad $A_i$,\quad the Hamilton---Jacobi equation has the form:
 \begin{equation}\label{1}
   g^{ij}P_iP_j=\lambda.
  \end{equation}
$$
\quad P_i=p_i+A_i,\quad p_i=\varphi_{,i}=\partial\varphi/\partial u^i, \quad \lambda = const.
$$
Let us consider the Klein---Gordon---Fock equation too:

$$\label{11}\hat{H}\varphi=(g^{ij}\hat{P}_i\hat{P}_j)\varphi = \lambda\varphi,\quad  \hat{P}_j = -\imath \hat{\nabla}_i + A_i.
$$
$\hat{\nabla}_i$ --- operator of the covariant derivative, with~metric-compatible connectivity, corresponding to the operator of the partial derivative ---  $\hat{\partial}_i = \imath \hat{p}_i$ \quad  with respect to the coordinate \quad  $u_i$; \quad $\varphi$ \quad is a field of a scalar particle with mass  \quad  $m=\sqrt{|\lambda|}.\quad$
For both equations, integrals of motion linear in momenta have the same form:
\begin{equation}\label{2}
\hat{Y}_A=\xi_A^iP_i+\gamma_A.
\end{equation}
It was proved in the articles \cite{Mag1} and \cite{OVV1}, that the conditions for the existence of an admissible electromagnetic field for the Klein-Gordon-Fock equation have exactly the same form as for the Hamilton---Jacobi equation. Therefore, we will consider only the Hamilton---Jacobi equation. All the results obtained in this case are also valid for the Klein---Gordon---Fock equation.

 The equation \ref{1} admits the integral of motion \ref{2}
 if and only if the functions \quad $\hat{H}=g^{ij}P_iP_j$ \quad and \quad
 $\hat{Y_A}$\quad commute with respect to the Poisson brackets:
 \begin{equation}\label{3} [\hat{H},\hat{Y}_A]_\mathcal{P}= (g^{il}{\xi^{j}_A}_{,l}+g^{jl}{\xi^{i}_A}_{,l}-g^{ij}_{,l}\xi_A^l)p_i p_j + 2g^{il}(\xi^{j}_A F_{ij}+\gamma_{A,i})p_l=0
 \end{equation}
 $$
  (F_{ij}=A_{j,i}-A_{i,j}).
  $$
 The ratios \ref{3} must be satisfied for any momentum value. Equating to zero the coefficients in front of the \quad $p_i$ \quad powers, one obtains:
 \begin{equation}\label{4}
 g^{il}{\xi^{j}_A}_{,l}+g^{jl}{\xi^{i}_A}_{,l}-g^{ij}_{,l}\xi_A^l=0, \end{equation} \begin{equation}\label{5}
 \gamma_{A,i} = \xi^{j}_A F_{ji}.
 \end{equation}
 It follows from the equations \ref{4} that the functions \quad$\xi ^{j}_A$ \quad are the components of the Killing vector.
 These functions form a groups of motions \quad $G_R$ \quad of \quad $rank (G_R)=r$.\quad
Unlike the free Hamilton-Jacobi equation, equation \ref{1}, generally speaking, has no integrals of motion. The system of equations \ref{5} specifies the conditions under which
integrals of motion of the form \ref{2} exist. This system was first developed by \cite{Mag1}.
 If the integrals of motion \ref{2} form an algebra with respect to Poisson brackets with the same structure constants as for the  algebra of the group operators, the following condition is satisfied:
  \begin{equation}\label{6}
  [\hat{X}_A,\hat{X}_B]_\mathcal{P}=C^D_{AB}\hat{X_D},
 \end{equation}
 In \cite{OVV1} it is stated that in an admissible electromagnetic field, the integrals of motion have the form:

 \begin{equation}\label{7}
 \hat{X}_A=\xi_A^i{p_i}.
 \end{equation}
 Let us prove this statement in the following theorem.

 \quad

 \begin{theor}
 {An admissible electromagnetic field does not deform the algebra of integrals of motion for the Hamilton-Jacobi and Klein---Gordon---Fock equations. The algebra operators for a free test particle and for a charged one have the form \ref{7}.}
 \end{theor}

 \quad

 \begin{proof}

  {As \quad $rank(G_r)=rank||\xi^i_A||=r$, \quad the transitivity subspace for the group motions \quad $G_R$ \quad is the $r$-dimensional space \quad $V_r$.\quad We represent the dependent operators of the group as follows:
\begin{equation}\label{8}
 \mathbf{X}_a = \omega^\alpha_a \mathbf{X}_\alpha,
\end{equation}
 Here \quad $\omega^\alpha_a$ \quad functions expressed in terms of the components of the Killing vectors \quad $\xi^i_A$\quad as follows:
 \begin{equation}\label{9}
 \omega^\alpha_a = \lambda^\alpha_\beta \xi^\beta_a, \quad \lambda^\alpha_\beta
 \xi^\beta_\gamma=\delta^\alpha_\gamma, \quad \xi^i_A=\delta^i_\beta\xi^\beta_A.
 \end{equation}
Transform the functions \quad $\gamma_\alpha$ \quad as follows:
$$
\gamma_\alpha = \xi_\alpha^i(\omega_i - A_i).
$$
From here it follows:
\begin{equation}\label{10}
 \hat{X}_\alpha=\xi_\alpha^i(p_i + \omega_i).
\end{equation}
Let us show that from the conditions \ref{6} it follows:
\begin{equation}\label{11}
\gamma_{\alpha}=\xi_\alpha^i A_i=0 \rightarrow \omega_\alpha =0,
\end{equation}
Substitute \ref{10} into \ref{6}. As a result, we get:
\begin{equation}\label{12}
[Y_\alpha,Y_\beta]_\mathcal{P}=C^\gamma_{\alpha\beta}\xi_\gamma^ip_i + \omega_{\beta|\alpha}-\omega_{\alpha|\beta} = C^\gamma_{\alpha\beta}\xi_\gamma^i(p_i+\omega_i) \quad (|_A =\xi_A^i\partial_i. \quad \omega_\alpha = \xi_\alpha^i \omega_i).
\end{equation}
Thus, the functions $\omega_\alpha$ obey the equations:
$$
 \omega_{\beta|\alpha}-\omega_{\alpha|\beta} = C^\gamma_{\alpha\beta}\omega_\gamma\quad \rightarrow \quad \omega_{i,j}=\omega_{j,i}\quad \rightarrow \quad \omega_i=\omega_{,i}.
$$
By the gradient transformation of the potential, the function \quad $\omega_i$\quad can be turned to \quad 0.\quad The theorem is proved.}
\end{proof}

\quad

Thereby, the condition for the admissible electromagnetic field existence has the form:

$$
(A_j\xi^j_A)_{,i} = \xi^{j}_A F_{ij}.
$$
Let us introduce a nonholonomic frame associated with the group \quad $G_R$:\quad
$$
\hat{\sigma}_j= \sigma^i_j\hat{\partial}_i = \delta^p_{j}\hat{\partial}_p+\delta^\alpha_{j}\xi_\alpha^\beta\hat{\partial}_\beta, \quad \hat{e}^j= e^j_i \hat{du}^i =\delta_p^{j}\hat{du}^p+\delta_\alpha^{j}\lambda^\alpha_\beta\hat{du}^\beta,
$$
non-holonomic components of the vector potential:
$$
\mathbf{A}_i=\sigma_i^jA_j,
$$
and also the potential projections onto the vector fields \quad $\xi^i_A$: \quad $\mathbb{A}_A=\xi^i_A A_i.$\quad Then the equation \ref{5}  takes the following form:

\begin{equation}\label{13}
\mathbb{A}_{A,i}=\xi^{j}_A F_{ij}.
\end{equation}

Consider two cases separately: \quad 1. $i=\alpha $ \quad and \quad 2. $i=p$.\quad

1.\quad $i=\alpha.$ \quad The system {\eqref{13}} can be reduced to the form:
$$
\mathbb{A}_{A|B} = \xi^{j}_A \xi^{i}_B(A_{j,i}-A_{i,j})=\mathbb{A}_{A|B}-\mathbb{A}_{B|A} + C^D_{AB}\mathbb{A}_D.
$$
From here it follows:
\begin{equation}\label{14}
\mathbb{A}_{B|A}=C^D_{AB}\mathbb{A}_D.
\end{equation}

2.\quad $i=p.$ \quad The system {\eqref{13}} can be reduced to the following form:
$$
(\xi^{j}_{B} A_{j})_{,p} = \xi^{j}_B (A_{j,p}-A_{p,j})= (\xi^{j}_{B} A_{j})_{,p} - A_{p|B} - \xi^{j}_{B,p} A_{j}.
$$
From here it follows:
\begin{equation}\label{15}
A_{p|A}=-\xi^i_{A,p}A_i,
\end{equation}
    The Systems \ref{14}, \ref{15} need to be examined for compatibility. Let us consider separately the cases when the order of the group \quad $G_R$ \quad coincides with the dimension of the transitivity subspace \quad $V_r$ \quad (R=r),\quad and when  \quad $R >r$.

 \subsection{Compatibility conditions for the case, when the order of the group \boldmath{$G_R$} \quad coincides with the dimension of the transitivity subspace  $V_r$.}

 Let us prove the compatibility of the system \eqref{14}, \eqref{15}. To do this, show that
 \begin{equation}\label{16}
 \mathbf{A}_{\beta|\alpha\gamma}-\mathbf{A}_{\beta|\gamma\alpha}= C^\sigma_{\alpha\beta}\mathbf{A}_{\sigma|\gamma}-C^\sigma_{\gamma\beta}\mathbf{A}_{\sigma|\alpha}.
 \end{equation}
 Since
 $$
 \mathbf{A}_{\beta|\alpha\gamma}-\mathbf{A}_{\beta|\gamma\alpha}= C^\sigma_{\gamma\alpha}\mathbf{A}_{\alpha|\sigma},
 $$
 $$
 C^\sigma_{\alpha\beta}\mathbf{A}_{\sigma|\gamma}-C^\sigma_{\gamma\beta}\mathbf{A}_{\sigma|\alpha}= (C^\sigma_{\alpha\beta}C^\rho_{\gamma\sigma}-C^\sigma_{\gamma\beta}C^\rho_{\alpha\sigma})\mathbf{A}_{\rho}, $$
 the system \eqref{16} is reduced to the following form:
 $$
 (C^\gamma_{\sigma\rho}C^\alpha_{\beta\gamma} + C^\gamma_{\rho\beta}C^\alpha_{\sigma\gamma} +C^\gamma_{\beta\sigma}C^\alpha_{\rho\gamma})\textbf{A}_\alpha = 0,
 $$
 which is fulfilled by the Bianchi identities.

 To prove the compatibility of the \eqref{15} system, it is necessary to show that
 \begin{equation}\label{17}
 A_{p|\beta\alpha}-A_{p|\alpha\beta}= C^\gamma_{\alpha\beta}A_{p|\gamma}= (\xi^\sigma_{\alpha ,p}\lambda^\gamma_\sigma\mathbf{A}_\gamma)_{|\beta} -(\xi^\sigma_{\beta,p}\lambda^\gamma_\sigma\mathbf{A}_\gamma)_{|\alpha}. \end{equation}
 Using the systems of equations \eqref{14}, \eqref{15}, as well as the consequences of the Killing equations:
 $$
 \xi^\gamma_{\alpha,p|\beta} -\xi^\gamma_{\beta,p|\alpha}= \xi^\nu_{\alpha,p}{\xi}^\gamma_{\beta,\nu} -\xi^\nu_{\beta,p}{\xi}^\gamma_{\alpha,\nu}+ C^\nu_{\beta\alpha}\xi^\gamma_{\nu,p}.
 $$
 we bring the conditions \eqref{17} to the form:
 \begin{equation}\label{18}
 \xi^\sigma_{\beta,p}(\xi^\rho_{\alpha,\sigma}\lambda^\gamma_{\rho}+ \lambda^\gamma_{\sigma|\alpha} + \lambda^\rho_{\sigma}C^\gamma_{\alpha\rho})\mathbf{A}_\gamma=\xi^\sigma_{\beta,p}(\xi^\rho_{\beta,\sigma}\lambda^\gamma_{\rho}+ \lambda^\gamma_{\sigma|\beta} + \lambda^\rho_{\sigma}C^\gamma_{\beta\rho})\mathbf{A}_\gamma
 \end{equation}
 It is easy to show that
 $$\xi^\rho_{\alpha,\sigma}\lambda^\gamma_{\rho}+ \lambda^\gamma_{\sigma|\alpha} + \lambda^\rho_{\sigma}C^\gamma_{\alpha\rho}=0.$$
 The compatibility of the system of equations \eqref{15} is proved.

\subsection{Compatibility conditions for the case when the order of the group  \boldmath{$G_R$} is greater than the dimension of the transitivity subspace  $V_r$.}
When studying the compatibility conditions for the systems of equations \eqref{14} \eqref{15}, one cannot use the results obtained in the previous section, since in this case the systems are overflowing, because the functions \quad $\mathbb{A}_B$\quad are linearly dependent with coefficients expressed in terms of the functions \quad $\xi^A_B$.\quad
Since the order of the group is greater than the dimension of the transitivity subspace, \quad $(R-r)$ \quad operators of the group are linear combinations of $r$ of the basic operators with variable coefficients. Without loss of generality, we can assume that \quad $rank{||\xi^i_\alpha||}=r \quad \rightarrow$ \quad the first $r$ of the $\hat{X}_\alpha$ operators are basic. In this case \quad $det||\xi^\alpha_\beta|| \ne 0$,\quad and the inverse matrix \quad $||\lambda^\alpha_\beta||$ \quad exists: $$\xi^\alpha_\gamma\lambda^\gamma_\beta=\delta^\alpha_\beta.$$
Then:
\begin{equation}\label{18}
\hat{X}_a=\xi^\beta_a \lambda^\alpha_\beta \hat{X}_\alpha \quad\rightarrow \quad \mathbb{A}_a=\xi^\beta_a \lambda^\alpha_\beta\mathbf{A}_\beta \quad \rightarrow \quad \mathbb{A}_a=\omega^\alpha_a\mathbf{A}_\alpha.
\end{equation}
Introduce the functions:
$$ \tilde{C}^\gamma_{AB}=C^\gamma_{AB}+\omega^\gamma_a C^a_{AB}.
$$
Then from the commutation relations:
$$[\hat{X}_A,\hat{X}_B]_P=C^D_{AB}\hat{X}_D,$$
one obtains the following identities:

\begin{equation}\label{19}
\omega^\gamma_{a|\alpha}=\tilde{C}^\gamma_{\alpha a}+\omega^\beta_a\tilde{C}^\gamma_{\beta\alpha},\quad \tilde{C}^\gamma_{ab}=\omega^\alpha_a\tilde{C}^\gamma_{\alpha b} -\omega^\alpha_b\tilde{C}^\gamma_{\alpha a}.
\end{equation}
The conditions \eqref{6} are reduced to the following systems of equations: \begin{equation}\label{20}
\mathbb{A}_{A|B} = \tilde{C}^\gamma_{B A}\mathbf{A}_\gamma
\end{equation}
\begin{equation}\label{21}
A_{p|A}=-\xi^\beta_{A,p}A_\beta=-\xi^\beta_{A,p}\lambda_\beta^\alpha \mathbf{A} _\alpha.
\end{equation}
Let us prove that the functions \quad $\mathbf{A}_\alpha$ \quad form a system of linearly dependent functions with coefficients that are rational functions of the components \quad $\xi^i_\alpha$ \quad and structural constants.
Indeed, the systems of equations \eqref{20}, \eqref{20}, depending on the values of the indices, can be broken down as follows:
\begin{equation}\label{22}
\mathbf{A}_{\beta|\alpha} = \tilde{C}^\gamma_{\alpha\beta}\mathbf{A}_\gamma;
\end{equation}
\begin{equation}\label{23}
\mathbf{A}_{\beta|a} = \tilde{C}^\gamma_{a\beta}\mathbf{A}_\gamma;
\end{equation}
\begin{equation}\label{24}
\mathbf{A}_{b|A} = \tilde{C}^\gamma_{A b}\mathbf{A}_\gamma;
\end{equation}
\begin{equation}\label{25}
\mathbf{A}_{p|a}= \omega^\alpha_{a}\mathbf{A}_{p,\alpha}=-\xi^\beta_{a,p}\lambda^\gamma_\beta \mathbf{A}_\gamma ;
\end{equation}
\begin{equation}\label{26}
\mathbf{A}_{p|\alpha} = -\xi^\beta_{\alpha,p}\lambda^\gamma_\beta \mathbf{A}_\gamma .
\end{equation}
From \eqref{23} it follows:
\begin{equation}\label{27}
(\tilde{C}^\gamma_{\alpha a}+\omega^\beta_a\tilde{C}^\gamma_{\beta\gamma})\mathbf{A}_\gamma=0 ,
\end{equation}
 Taking into account \eqref{19} one obtains:
\begin{equation}\label{28}
\omega^\gamma_{a|\alpha}\mathbf{A}_\gamma=0.
\end{equation}

Since
$$\xi^\alpha_a=\omega^\beta_a\xi^\alpha_\beta,$$
from \eqref{25},\eqref{26} it follows:
\begin{equation}\label{29}
\omega^\gamma_{a|p}\mathbf{A}_\gamma=0 \rightarrow \omega^\gamma_{a,i}\mathbf{A}_\gamma=0. \end{equation}
Obviously, \quad $\omega^\gamma_{a,i} \ne 0,$ \quad otherwise \quad $\omega^\gamma_{a}=const.$ \quad and \quad $G_R$ \quad is reduced to \quad $G_r$. The statement is proven true.

The system \eqref{24} is satisfied identically due to the conditions \eqref{19}.
Consider the \eqref{26} system. The compatibility conditions are as follows:
\begin{equation}\label{30}
\mathbf{A}_{p|\alpha\gamma}-\mathbf{A}_{p|\gamma\alpha} =(\xi^\beta_{\gamma,p}A_\beta)_{|\alpha}-(\xi^\beta_{\alpha,p}A_\beta)_{|\gamma}.
\end{equation}
Using the condition \eqref{22}, as well as the consequence from the Killing equations:
$$(\xi^\gamma_{\alpha|\beta}-\xi^\gamma_{\beta|\alpha})_{,p}=\omega^\nu_{a|p}\xi^\gamma_\nu \tilde{C}^a_{\alpha\beta}+\xi^\gamma_{\nu,p} \tilde{C}^\gamma_{\alpha\beta} $$
we get:
\begin{equation}\label{31}
\omega^\gamma_a\lambda^\mu_\nu(\xi^\nu_{\beta,p}C^a_{\alpha\mu}- \xi^\nu_{\alpha,p}C^a_{\beta\mu})\mathbf{A}_\gamma=0.
\end{equation}
Let us supplement this system with equations \eqref{11},\eqref{12}:
\begin{equation}\label{32}
\omega^\gamma_{a,i}\mathbf{A}_\gamma=0.
\end{equation}
Let $K$ be the number of independent equations in the system \eqref{31}, \eqref{32}, and the numbers $\mu$ range from 1 to $K.$ Let us randomly enumerate all pairs (in \eqref{32}) and all triples (in \eqref{31}) of subscripts in the coefficients before \quad $\mathbf{A}_\alpha$ \quad and represent these coefficients as elements of the matrix \quad $\hat{W}(0)=||{W{(0)}}_\mu^\alpha||$. \quad Thus, the functions \quad $\mathbf{A}_\alpha$ \quad satisfy the system of equations:
\begin{equation}\label{33}
 {W{(0)}}_\mu^\alpha\mathbf{A}_\alpha=0.
\end{equation}
Denote:
$$
\tilde{r} = r- Rank||\hat{W}{(0)}||\quad (r-\tilde{r} \ne 0),
$$

In the matrix \quad $\hat{W}{(0)}$\quad choose an arbitrary minor \quad$\hat{Z}$ \quad of  $rank =\tilde{r}$.\quad The indices numbering the rows of this minor are denoted by the letters \quad $ \tilde{a}, \tilde{b},
\tilde{c}.$ \quad The rest of the minor columns will be denoted by the letters $\tilde{\alpha},\tilde{\beta},\tilde{\gamma}.$
These letters will also be used to number the columns that form the non-singular square matrix \quad$\hat{\Omega}$ \quad of\quad $rank=\hat{r}$ \quad in this minor.

Then the solution of the system of equations \eqref{22} can be represented as
\begin{equation}\label{33a}
\mathbf{A}_{\tilde{a}}=\theta^{\tilde{\alpha}}_{\tilde{a}}\mathbf{A}_{\tilde{\alpha}},
\end{equation}
where
\begin{equation}\label{34}
\theta^{\tilde{\alpha}}_{\tilde{a}}=
-||\hat{\Omega}||^{\tilde{b}}_{\tilde{a}} Z^{\hat{\alpha}}_{\tilde{b}}.
\end{equation}
The functions \eqref{34} convert the systems of equations \eqref{31} and \eqref{32} into identities: $$
\omega^{\tilde{\gamma}}_{a|i} + \omega^{\tilde{a}}_{a|i}\theta^{\tilde{\gamma}}_{\tilde{a}}=0 \quad \quad (\omega^{\tilde{\gamma}}_{a} + \omega^{\tilde{a}}_{a}\theta^{\tilde{\gamma}}_{\tilde{a}})\lambda^\mu_\nu(\xi^\nu_{\beta,p}C^a_{\alpha\mu}- \xi^\nu_{\alpha,p}C^a_{\beta\mu})=0.
$$
Let's go back to the system of equations \eqref{22}. Substitute expressions \eqref{33} into it. As a result, we get the following subsystems:
\begin{equation}\label{35}
\mathbf{A}_{\tilde{\alpha}|\alpha} = \tilde{C}^\gamma_{\alpha\tilde{\alpha}}\mathbf{A}_\gamma =(\tilde{C}^{\tilde{\gamma}}_{\alpha\tilde{\alpha}} + \theta^{\tilde{\gamma}}_{\tilde{b}}\tilde{C}^{\tilde{b}}_{\alpha\tilde{\alpha}})\mathbf{A}_{\tilde{\gamma}}. \end{equation}
\begin{equation}\label{36}
\mathbf{A}_{\tilde{a}|\alpha} = \tilde{C}^\gamma_{\alpha\tilde{a}}\mathbf{A}_\gamma =(\tilde{C}^{\tilde{\gamma}}_{\alpha\tilde{a}} + \theta^{\tilde{\gamma}}_{\tilde{b}}\tilde{C}^{\tilde{b}}_{\alpha\tilde{a}})\mathbf{A}_{\tilde{\gamma}}. \end{equation}
Using \eqref{33} and \eqref{35}, the system of equations \eqref{36} can be represented as a system of algebraic equations:
\begin{equation}\label{37}
\quad
(\theta^{\tilde{\gamma}}_{\tilde{a}|\alpha} + \theta^{\tilde{\beta}}_{\tilde{a}}(\tilde{C}^{\tilde{\gamma}}_{\alpha\tilde{\beta}} + \theta^{\tilde{\gamma}}_{\tilde{b}}\tilde{C}^{\tilde{b}}_{\alpha\tilde{\beta}})-(\tilde{C}^{\tilde{\gamma}}_{\alpha\tilde{a}} + \theta^{\tilde{\gamma}}_{\tilde{b}}\tilde{C}^{\tilde{b}}_{\alpha\tilde{a}}))\mathbf{A}_{\tilde{\gamma}}=0 \end{equation}

Combining \eqref{37} and the compatibility conditions for the systems of equations, \eqref{35}-- \eqref{26}, one obtains new compatibility conditions in the form similar to the conditions \eqref{33}:
$$
{W{(1)}}_{\mu_1}^{\tilde{\alpha}}\mathbf{A}_{\tilde{\alpha}}=0.
$$
The index \quad $\mu_1$ \quad is constructed in the same way as the index in the matrix \quad  ${W{(0)}}_\mu ^\alpha$.

If \quad ${W{(1)}}_{\mu_1} ^{\tilde{\alpha}}=0,$\quad the solution has the form \eqref{33a} and the systems of equations \eqref{13}, \eqref{15} are compatible.

If \quad ${W{(1)}}_{\mu_1} ^{\tilde{\alpha}}\ne 0$, \quad this procedure should be repeated until at step ($\varkappa$) the condition is met \quad ${W{(\varkappa)}}_{\mu_\varkappa} ^{\tilde{\alpha}_\varkappa}=0$ ,\quad and the remaining system of differential equations:

\begin{equation}\label{38}
\mathbf{A}_{{\tilde{a}_\varkappa}|\tilde{\alpha}} = \tilde{C}^{\tilde{\gamma}_\varkappa}_{\tilde{\alpha}({{\tilde{a}_\varkappa}})}\mathbf{A}_{\gamma_\varkappa}, \quad \mathbf{A}_{\tilde{\alpha}} = \vartheta_{\tilde{\alpha}}^{\tilde{\alpha}_\varkappa}\mathbf{A}_{({\tilde{a}_\varkappa})}
\end{equation}
will  be compatible.
Otherwise, the only solution to the system \eqref{22} will be - $A_i= 0$. Note that for a specific space on which the group \quad $G_R$ \quad acts,\quad $\omega^{\alpha}_{a}, \quad \theta^{\tilde{\gamma}}_{\tilde{a}},\quad \vartheta_\alpha^{(\tilde{\alpha}_N)}$, \quad  are the given functions. Therefore, checking the compatibility conditions does not imply the solution to any equations.

\section{Admissible electromagnetic fields for the groups of motions  \boldmath{$G_4$} acting on transitivity subspaces of a spacetime manifold}

 The complete classification of  admissible electromagnetic field potential is an obvious continuation of the classification problem, solved by A.Z. Petrov (\cite{Petrov}). Despite its obvious importance, the first papers (\cite{Mag1}, \cite{Mag2}, \cite{OVV1}, \cite{OVVA}) on this topic appeared quite recently.  This section continues the research begun in these articles. The potentials of all admissible electromagnetic fields are found for the case when the group of motions \quad $ G_4 $ \quad acts nontransitively on the space \quad $ V_4 $. \quad The results obtained in the previous section are used.

Let us clarify the notation of the indices used in this section. As is known, every four-parameter Lie group \quad $ G_4 $ \quad has a three-parameter subgroup \quad $ G_3 $ \quad (see \cite{Petrov}). Therefore, as the basic operators \quad $\hat{X}_\alpha$ \quad of the group \quad $G_4$,\quad one can choose the operators of the subgroup \quad $G_3$.\quad The indices introduced earlier in this section vary within the following limits: \quad $i, j = 0, \dots 3$;\quad $\alpha, \beta, \gamma, =1, \dots, 3;$\quad $ A, B = 1, \dots 4; \quad $ $a = 4; $ \quad
The following algorithm is used for constructing the potentials of an admissible electromagnetic field.

1. Using the known Killing vectors \quad $\xi^\alpha_A$ \quad of the group \quad $G_4$\quad there can be found the matrix \quad$||\lambda^\alpha_\beta||$\quad and the functions \quad $\omega^\alpha=-\lambda^\alpha_\beta \xi^\beta_4$.

2. The functions \quad $\tilde{C}^\alpha_{AB}$ \quad are calculated and the matrix
$$ ||W^\gamma_{\alpha}(0)||=||\tilde{C}^\gamma_{\alpha 4}+\omega^\beta\tilde{C}^\gamma_{\beta\alpha}||
$$ is constructed using the \eqref{19} formula.

3. The solution to the system of equations is found
\begin{equation}\label{40}
W^\gamma_{\alpha}(0)\mathbf{A_\gamma}=0
\end{equation} in the following form:
$$ \mathbf{A}_{\tilde{a}}=\theta^{\tilde{\alpha}}_{\tilde{a}}\mathbf{A}_{\tilde{\alpha}}. \quad rank ||\theta^{\tilde{\alpha}}_{\tilde{a}}||=(3-\tilde{r}).
$$

4. The component \quad $\mathbf{A_4}$ \quad is calculated by the formula:
$$
\mathbf{A}_a = \mathbf{A}_4= \omega^\alpha\mathbf{A}_\alpha,
$$
the system \eqref{11} compatibility is checked and the solutions to the remaining equations
\begin{equation}\label{41} \mathbf{A}_{\tilde{\alpha}|\tilde{\beta}}=\tilde{C}^{\tilde{\gamma}}_{\tilde{\beta}\tilde{\alpha}}\mathbf{A}_{\tilde{\gamma}}. \end{equation}
are found.

\quad

 As already noted, group operators, spacetime metrics and canonical coordinate systems for
 all groups of motions \quad $G_R$ \quad  were found in \cite{Petrov}.
 Below, for ease of use, these formulas and structural constants are given for each considered
 group.  Moreover in every subsection the results of calculation  are presented in the following
 order. Matrices $\hat{\lambda}$, \quad $\hat{W}$,\quad
are given as well as nonholonomic \quad $\mathbf{A}_A$ \quad  and holonomic \quad $A_ i$\quad  components of the potentials of the admissible electromagnetic field. The functions denoted by the letter
 \quad $a$ \quad with a single lower index depend only on the variable \quad $u^0$, \quad $(e_0)^2 = 1$

 The transitivity hypersurface of spacetime \quad $V_4$ \quad in the canonical coordinate system is given by the equation:
$$
\phi(u^i) = u^0 = const.
$$
If  \quad $g^{ij} \phi_{,i} \phi_{,j} \ne  0 $,\quad
the surface is called non-isotropic (non-null) and is denoted by \quad $V_3$. \quad Otherwise
the hypersurface is called isotropic (null)  and is denoted by \quad $ V^*_3 $.

Each case when the group \quad $ G_4 $ \quad acts transitively on the hypersurfaces
\quad $ V_3 $ \quad and  \quad $ V^*_3 $, \quad also  when the mentioned above subgroups
\quad $ G_3 $ \quad act transitively on \quad $ V_2 $ \quad and \quad $ V^*_2 $ \quad
(they are the subspaces of the hypersurface  \quad $ V_3 $) \quad is considered separately.

The notation  $G_R(N)$ means that this group has an order $R$ and the group structure is of the type $N$ according to the Bianchi classification. \cite{Bianchi}

\subsection{The groups \boldmath{$G_4$} act transitively on a non null subspace $V_3$}

 When the group \quad $G_4$ \quad acts transitively on a non null subspace \quad $V_r,\quad r\leq3$ \quad it follows from \cite{Petrov} that \quad $(\xi^\alpha_A)_{,0}=0.$  \quad According to \eqref{26} this means: \quad $ A_0=A_0(u^0)$, \quad
which is equivalent to
$$
A_0=0.\quad
$$

\subsubsection{Group $G_4 (I)$}.

Metric, in which this group acts, has the form:
$$
ds^2=2a_1 du^1 du^3 + a_2( du^2 + u^1 du^3)^2 +e_0(du^0)^2.
$$

Let us present operators of the group:
$$
X_1 =p_1,\quad X_2 = p_2,\quad X_3 =u^2 p_1-p_3,\quad X_4 =u^2 p_2- u^3 p_3,
$$

and structural constants:
$$
 C^\gamma_{A1}=0,\quad C^\gamma_{23}=\delta^\gamma_1, \quad C^\gamma_{24} = \delta^\gamma_2, \quad C^\gamma_{34}=-\delta^\gamma_3. $$
Matrix \quad $\hat{\lambda}$, \quad $\hat{W}(0)$  \quad and functions  \quad $\omega^\alpha$, \quad which defined by the formulas \eqref{9}, \eqref{33}, have the form:
$$
||\lambda^\alpha_\beta||=
\begin{pmatrix}1 & 0 & 0 \\
   0 & 1 & 0 \\
     u^2& 0 & -1
\end{pmatrix},\quad
$$

$$
\omega^\alpha=-u^2 u^3\delta^\alpha_1 + u^2\delta^\alpha_2
+ u^3\delta^\alpha_3
$$

Matrix:
$$
||W^\alpha_\beta||=
\begin{pmatrix}0 & 0& 0&\\u^3 & -1 & 0 \\
    -u^2 & 0 & 1
\end{pmatrix},\quad
$$

From the system of equations \eqref{40} it follows:
\begin{equation}\label{42} \mathbf{A_2}= u^3\mathbf{A_1}, \quad \mathbf{A_3} = u^2\mathbf{A_1}.
\end{equation}

Substitute the resulting expressions into \eqref{41}. As a result, we get:
$$
\mathbf{A}_{1,2}=\mathbf{A}_{2,2}=\mathbf{A}_{\alpha,1}=\mathbf{A}_{1,3}=\mathbf{A}_{3,3}=0;
$$
$$\mathbf{A}_{3,2}=\mathbf{A}_{1}, \quad \mathbf{A}_{2,3} = \mathbf{A}_{1}.
$$

The solution is:
$$ \mathbf{A}_{1}=a_0, \quad \mathbf{A}_{2} =a_0 u^3,\quad \mathbf{A}_{3}=a_0u^2.$$

The components of the potential of the admissible electromagnetic field are found by the formula \begin{equation}\label{43}
A_\alpha=\mathbf{A}_\beta \lambda^\beta_\alpha, \quad A_0=0
\end{equation}
and have the form:
$$ \quad A_0=0, \quad \quad A_1=a_0, \quad A_2=a_0u^3, \quad A_3=0. \quad
$$

\subsubsection{Group $G_4(III)$.}

Metric:
$$
ds^2=a_1((du^1)^2 + (du^3)^2) + a_2( du^2 + u^1  du^3)^2 +e_0(du^0)^2.
$$

Group operators:
$$
X_1 =p_1,\quad   X_2 = p_2,\quad  X_3 =u^2 p_1+p_3,\quad X_4 =\frac{(u^2)^2+(u^3)^2}{2}p_1 + u^3 p_2 + u^2 p_3.
$$

Structural constants:
$$
 C^\gamma_{23}=\delta^\gamma_1, \quad  C^\gamma_{24} = \delta^\gamma_3, \quad  C^\gamma_{34}=-\delta^\gamma_2.
$$

Matrix $\hat{\lambda}$:
$$
||\lambda^\alpha_\beta||=
\begin{pmatrix}1 & 0 & 0 \\
   0 & 1  & 0  \\-u^2 & 0 & 1
\end{pmatrix},\quad
$$

Functions $\omega^\alpha$:
$$
\omega^\alpha=\frac{(u^2)^2 + (u^2)^2}{2}\delta^\alpha_1 + u^3\delta^\alpha_2
+ u^2\delta^\alpha_3
$$

Matrix $W^\alpha_\beta$:
$$
||W^\alpha_\beta||=
-\begin{pmatrix}0 & 0& 0&\\-u^2 & 0 & 1 \\
   u^3 & 1  & 0 \end{pmatrix}.
$$

From the system of equations \eqref{40} it follows:
\begin{equation}\label{42}
\mathbf{A_2}=-u^3\mathbf{A_1}, \quad \mathbf{A_3} = u^2\mathbf{A_1}.
\end{equation}

Substitute the resulting expressions into \eqref{41}. As a result, we get:
$$
\mathbf{A}_{1,\alpha}= 0, \quad \mathbf{A}_{2,2}=\mathbf{A}_{2,1}=\mathbf{A}_{3,1}=\mathbf{A}_{3,3}=0;
$$
$$\mathbf{A}_{3,2}=\mathbf{A}_{1}, \quad  \mathbf{A}_{2,3} = -\mathbf{A}_{1}.
$$

The solution is:
$$ \mathbf{A}_{1}=a_0, \quad \mathbf{A}_{2} =-a_0 u^3,\quad \mathbf{A}_{3}=a_0u^2.$$

The components of the potential of the admissible electromagnetic field are found by the formula \eqref{43} and have the form:
$$
\quad A_0=0, \quad \quad A_1=a_0, \quad A_2=-a_0u^3, \quad A_3=0. \quad
$$

\subsubsection{Group \boldmath{$G_4(IV)$}}.

Metric:
$$
ds^2=a_1(du^1)^2 +2a_2\exp{u^1} du^2 du^3 +e_0(du^0)^2.
$$

Group operators:
$$
X_1 =p_1,\quad   X_2 = p_2,\quad  X_3 =u^1 p_1+p_3,\quad X_4 = u^2 p_2 + p_3.
$$

Structural constants:
$$
 C^\gamma_{13}=\delta^\gamma_1, \quad  C^\gamma_{24} = \delta^\gamma_2.
$$

Matrix $\hat{\lambda}$:
$$
||\lambda^\alpha_\beta||=
\begin{pmatrix}1 & 0 & 0 \\
   0 & 1  & 0  \\-u^2 & 0 & 1
\end{pmatrix},\quad
$$

Functions $\omega^\alpha$:
$$
\omega^\alpha=-u^1 \delta^\alpha_1 + \delta^\alpha_3
$$

Matrix $W^\alpha_\beta$:
$$
W^\alpha_\beta=
\begin{pmatrix}1 & 0& 0&\\0 & -1 & 0 \\
   u^1 & 0  & 0 \end{pmatrix}.
$$

From the system of equations \eqref{40} it follows:
$$
\mathbf{A_1}=\mathbf{A_2}=0.
$$

From the system \eqref{41}. it follows:
$$
\mathbf{A}_{3,\alpha}=0 \quad \rightarrow \quad \mathbf{A}_{3} =a_0.
$$

The components of the potential of the admissible electromagnetic field are found by the formula \eqref{43} and have the form:
$$
\quad A_0= A_1= A_2=0, \quad A_3=a_0. \quad
$$

\subsubsection{Group \boldmath{$G_4(V)$}}.

Metric:
$$
ds^2=a_1(du^1)^2 +a_2( (du^2)^2 + (du^3)^2)\exp{2u^1} +e_0(du^0)^2.
$$

Group operators:
$$
X_1 =p_1,\quad   X_2 = p_2,\quad  X_3 =u^1 p_1 +u^2p_2-p_3,\quad X_4 = -u^2 p_1 +u^1 p_2.
$$

Structural constants:
$$
 C^\gamma_{13}=C^\gamma_{42}=\delta^\gamma_1, \quad C^\gamma_{23}=C^\gamma_{14}=\delta^\gamma_2.
$$

Matrix $\hat{\lambda}$:
$$
||\lambda^\alpha_\beta||=
\begin{pmatrix}1 & 0 & 0 \\
   0 & 1  & 0  \\u^1 & u^2 & -1
\end{pmatrix},\quad
$$

Functions $\omega^\alpha$:
$$
\omega^\alpha=-u^2 \delta^\alpha_1 + u^1\delta^\alpha_2.
$$

Matrix $W^\alpha_\beta$:
$$
W^\alpha_\beta=
\begin{pmatrix}0 & -1 & 0 & \\1 & 0 & 0 \\
   u^2 & -u^1  & 0 \end{pmatrix}.
$$

From the system of equations \eqref{40} it follows:
$$
\mathbf{A_1}=\mathbf{A_2}=0.
$$

Substitute into the system \eqref{41}. As a result, we get:
$$
\mathbf{A}_{3,\alpha}=0 \quad \rightarrow \quad \mathbf{A}_{3} =a_0.
$$

The components of the potential of the admissible electromagnetic field are as follows:
$$
\quad A_0= A_1= A_2=0, \quad A_3=a_0. \quad
$$

\subsubsection{Group \boldmath{$G_4(VII)$}}.

Metric:
$$
ds^2=4a_1du^1(du^2-u^2du^3) + a_2(du^2-u^2du^3)^2 - a_1(du^3)^2 +e_0(du^0)^2.
$$

Group operators:
$$
X_1 =p_1,\quad   X_2 = p_2 \exp{u^3},\quad  X_3 = p_3,\quad X_4 = (p_1 -(u^2)^2p_2-2u^2p_3)\exp{-u^3}.
$$

Structural constants:
$$
 C^\gamma_{1\alpha}=0, \quad C^\gamma_{32}=\delta^\gamma_2, \quad C^\gamma_{42}=2\delta^\gamma_3, \quad C^A_{43}=\delta^A_4 \quad \rightarrow
 \quad \tilde{C}^\gamma_{43}=\omega^\gamma.
$$

Matrix $\hat{\lambda}$:
$$
||\lambda^\alpha_\beta||=
\begin{pmatrix}1 & 0 & 0 \\
   0 & \exp{-u^3} & 0  \\0 & 0 & 1
\end{pmatrix},\quad
$$

Functions $\omega^\alpha$:
$$
\omega^\alpha=(\delta^\alpha_1-(u^2)^2\exp{(-u^3)} \delta^\alpha_2 - 2u^2\delta^\alpha_3)\exp{-u^3}.
$$

Matrix $W^\alpha_\beta$:
$$
W^\alpha_\beta=
\begin{pmatrix}0 & 0 & 0 & \\0 &u^2\exp{(-u^3)}& 1 \\ \exp{(-u^3)}&
-2(u^2\exp{(-u^3)})^2 &-2u^2\exp{(-u^3)} \end{pmatrix}.
$$

From the system of equations \eqref{40} it follows:
$$
\mathbf{A}_{\alpha,1}=0,\quad \mathbf{A}_{\alpha}=-u^2\exp{(-u^3)}\mathbf{A_2}.
$$

Substitute this into the system \eqref{41}. As a result, we get:
$$
\mathbf{A}_{\alpha,1} =0,\quad \mathbf{A}_{\alpha,2} =-\delta_{\alpha3}\exp{(-u^3)}\mathbf{A_2}\quad \mathbf{A}_{\alpha,3} = \mathbf{A}_{2}\delta_{\alpha2}.
$$
From here it follows:
$$
\mathbf{A}_{1}=0,\quad \mathbf{A}_{2} = a_0 \exp{u^3}, \quad \mathbf{A}_{3} =-a_0u^2.
$$

The components of the potential of the admissible electromagnetic field are as follows:
$$
\quad A_0= A_1=0, \quad  A_2=a_0, \quad A_3=-a_0 u^2. \quad
$$

\subsubsection{Group \boldmath{$G_4(IV)$}}.

Metric:
$$
ds^2=a_1((du^1)^2+\sin^2{u^1}(du^2)^2) + a_2(\cos{u^1}du^2+du^3)^2  +e_0(du^0)^2.
$$

Group operators:
$$
X_1 =p_1,\quad   X_2 = p_2,\quad  X_3 =-\frac{\sin{u^3} \sin{u^1}}{\cos{u^3}}p^1 + \frac{\sin{u^1}}{\sin{u^3}}p_2 + \cos{u^1} p_3,\quad X_4 = \frac{\partial{X_3}}{\partial{u^2}}.
$$

$$
X_1 =p_1,\quad   X_2 = p_2 ,\quad  X_3 =-\frac{\cos{u^3} \sin{u^1}}{\sin{u^3}}p^1 + \frac{\varepsilon\sin{u^1}}{\sin{u^3}}p_2 + \cos{u^1} p_3,\quad X_4 = \frac{\partial{X_3}}{\partial{u^2}}.
$$

Structural constants:
$$
 C^A_{13}=\delta^A_4  \rightarrow
  \tilde{C}^\gamma_{13}=\omega^\gamma, \quad C^A_{41}=\delta^A_3, \quad C^A_{34}=\delta^A_1.
$$

Matrix $\hat{\lambda}$:
$$
||\lambda^\alpha_\beta||=
\begin{pmatrix}1 & 0 & 0 \\
   0 & 1 & 0  \\\frac{\sin{u^1}\cos{u^3}}{\cos{u^1}\sin{u^3}}& -\frac{\sin{u^1}}{\cos{u^1}\sin{u^3}} &\frac{1}{\cos{u^1}}
\end{pmatrix},\quad
$$

Functions $\omega^\alpha$:
$$
\omega^\alpha=-\frac{\delta^\alpha_1\cos{u^3}}{\cos{u^1}\sin{u^3}}
+\frac{\delta^\alpha_2}{\cos{u^1}\sin{u^3}} - \frac{\delta^\alpha_3 \sin{u^1}}{\cos{u^1}}.
$$

Elements of the matrix $W^\alpha_\beta$:
$$
W^\alpha_\gamma=\delta_{\alpha1}(\delta^\gamma_3 + \omega^3\omega^\gamma)-
\delta_{\alpha3}(\delta^\gamma_1 + \omega^1\omega^\gamma).
$$

From the system of equations \eqref{40} it follows:
$$
\mathbf{A}_{3}=-\mathbf{A}_{\alpha}\omega^\alpha\omega^3,\quad \mathbf{A}_{1}=-\mathbf{A}_{\alpha}\omega^\alpha\omega^2.
$$
Denote: \quad
$$
\mathbf{A}=-\mathbf{A}_{\alpha}\omega^\alpha \quad \rightarrow \quad \mathbf{A}_{3}=\mathbf{A}\omega^3,\quad \mathbf{A}_{1}=\mathbf{A}\omega^2.
$$
From here it follows:

\begin{equation}\label{44}
\mathbf{A}=-\mathbf{A}_{2}\sin{u^3}\cos{u^1} \quad
\mathbf{A}_{1}=\mathbf{A}_2\cos{u^3},\quad \mathbf{A}_{3}=\mathbf{A}_2\sin{u^1}\sin{u^3}.
\end{equation}

 From the system \eqref{41} we get
$$
\mathbf{A}_{2,\alpha} = 0 \quad \rightarrow \quad \mathbf{A}_{2} = a_0.
$$
Thus, the components of the potential of the admissible electromagnetic field are as follows:
$$
A_{0}= A_{3}=0, \quad  A_{2}=a_0, \quad A_{1}=a_{0}\cos{u^3}.
$$

\subsection{The subgroups \boldmath{$G_3$} of the group $G_4$ act  transitively  on  subspce  $V_2$ \quad of \quad $V_3$ }

\subsubsection{$G_3$ is a subgroup  of the group $G_4(I)$}.
Metric:
$$
ds^2=a_1(u^3)^{\frac{1}{c-1}}(2du^1du^3+(du^2)^2) + a_2(\frac{du^3}{u^3})^2  +e_0(du^0)^2.
$$
Group operators:
$$
X_1 =p_1,\quad   X_2 = p_2 ,\quad  X_3 =2cu^1p_1 +u^2p_2+2(1-c)u^3p_3,
\quad X_4 = u^2p_1 -u^3p_2.
$$
Structural constants:
$$
 C^A_{13}=2c\delta^A_1\quad C^A_{23}=\delta^A_2, \quad C^A_{24}=\delta^A_1 C^A_{34}=(1-2c)\delta^A_4 \quad \rightarrow \quad
 \tilde{ C}^\alpha_{34}=(1-2c)(u^3\delta^\alpha_2 -u^2\delta^\alpha_1).
$$
Matrix $\hat{\lambda}$:
$$
||\lambda^\alpha_\beta||=
\begin{pmatrix}1 & 0 & 0 \\
   0 & 1 & 0  \\\frac{cu^1}{(c-1)u^3}& \frac{u^2}{2(c-1)u^3} &\frac{1}{2(1-c)u^3}.
\end{pmatrix},\quad
$$
Functions $\omega^\alpha$:
$$
\omega^\alpha=u^2\delta^\alpha_1-u^3\delta^\alpha_2.
$$
Matrix $\hat{W}$:
$$
W^\gamma_\alpha=-\delta^{\gamma}_1(\delta_{\alpha 2} + u^2 \delta_{\alpha3}) +2(1-c)\delta^{\gamma}_2 \delta_{\alpha3}.
$$
From the system of equations \eqref{40} it follows:
$$
\mathbf{A}_{1}=\mathbf{A}_{2} = 0.
$$
Taking this into account the \eqref{36} will take the form:
$$
\mathbf{A}_{3,\alpha}=0 \quad \rightarrow \quad \mathbf{A}_{3}=2(1-c)a_0.
$$
The components of the potential of the admissible electromagnetic field are as follows:
$$
A_{0}= A_{3}=A_{2}=0, \quad A_{1}=\frac{a_{0}}{u^3}
$$

\subsubsection{\boldmath{$G_3$} is a subgroup  of the group $G_4(II)$}.
Metric:
$$
ds^2=a_1(2du^1du^3+(du^2)^2)\exp{-2u^3} + a_2(u^3)^2  +e_0(du^0)^2.
$$
Group operators:
$$
X_1 =p_1,\quad   X_2 = p_2 ,\quad  X_3 =2u^1p_1 +u^2p_2+p_3,
\quad X_4 = u^2p_1 -u^3p_2.
$$
Structural constants:
$$
 C^A_{13}=2\delta^A_1,\quad C^A_{23}=\delta^A_2, \quad C^A_{24}=\delta^A_1,\quad C^A_{43}=\delta^A_2 +\delta^A_4 \quad \rightarrow \quad
 \tilde{C}^\alpha_{34}=(u^3-1))\delta^\alpha_2 -u^2\delta^\alpha_1.
$$
Matrix $\hat{\lambda}$:
$$
||\lambda^\alpha_\beta||=
\begin{pmatrix}1 & 0 & 0 \\
   0 & 1 & 0  \\-2u^1& -u^2 &1
\end{pmatrix}.\quad
$$
Functions $\omega^\alpha$:
$$
\omega^\alpha=u^2\delta^\alpha_1-u^3\delta^\alpha_2.
$$
Matrix $\hat{W}$:
$$
W^\gamma_\alpha=\delta^{\gamma}_2(-\delta_{\alpha 2} + \delta_{\alpha3}) -u_2\delta^{\gamma}_1 \delta_{\alpha3}.
$$
From the system of equations \eqref{40} it follows:
$$
\mathbf{A}_{1}=\mathbf{A}_{2} = 0.
$$
Taking this into account the \eqref{36} will take the form:
$$
\mathbf{A}_{3,\alpha}=0 \quad \rightarrow \quad \mathbf{A}_{3}=a_0.
$$
The components of the potential of the admissible electromagnetic field are as follows:
$$
A_{0}= A_{1}=A_{2}=0, \quad A_{3}=a_{0}.
$$
\subsubsection{\boldmath{$G_3$} is a subgroup  of the group $G_4(III)$}.

Metric:
$$
ds^2=R^{-1}_3(a_1S(2du^1du^3+(du^2)^2) + a_2(u^3)^2)  +e_0(du^0)^2,
$$
$$
\quad R_3=(u_3)^2 +2u^3\sin{c}+1, \quad S=\exp{((\sin{2c})\arctan\frac{u^3+\sin{c}}{\cos{c}})} .
$$
Group operators:
$$
X_1 =p_1,\quad   X_2 = p_2 ,\quad  X_3 =\frac{(4u^1\sin{c}+(u^2)^2)p_1}{2} -u^2u^3p_2-R_3p_3,
\quad X_4 = u^2p_1 -u^3p_2.
$$
Structural constants:
$$
 C^A_{13}=2\delta^A_1\sin{c},\quad C^A_{23}=\delta^A_1, \quad C^A_{24}=\delta^A_1,\quad C^A_{34}=\delta^A_2 -\delta^A_4\sin{c} \quad \rightarrow \quad
 \tilde{C}^\alpha_{34}=\delta^\alpha_2 -2(\delta^\alpha_1u^2-\delta^\alpha_2u^3)\sin{c}.
$$
Matrix $\hat{\lambda}$:
$$
||\lambda^\alpha_\beta||=
\begin{pmatrix}1 & 0 & 0 \\
   0 & 1 & 0  \\\frac{(4u^1\sin{c}+(u^2)^2)}{2R_3}&\frac{u^2u^3}{R_3} &-\frac{1}{R_3}
\end{pmatrix}.\quad
$$
Functions $\omega^\alpha$:
$$
\omega^\alpha=u^2\delta^\alpha_1-u^3\delta^\alpha_2.
$$
Matrix $\hat{W}$:
$$
W^\gamma_\alpha=\delta^{\gamma}_1(-\delta_{\alpha_2} + u^2u^3\delta_{\alpha3}) -R_3\delta^{\gamma}_2 \delta_{\alpha3}.
$$
From the system of equations \eqref{40} it follows:
$$
\mathbf{A}_{1}=\mathbf{A}_{3} = 0.
$$
Taking this into account the \eqref{36} will take the form:
$$
\mathbf{A}_{3,\alpha}=0 \quad \rightarrow \quad \mathbf{A}_{3}=-a_0.
$$
The components of the potential of the admissible electromagnetic field are as follows:
$$
A_{0}= A_{1}=A_{2}=0, \quad A_{3}=\frac{a_{0}}{R_3}.
$$

\subsubsection{\boldmath{$G_3$} are  subgroup  of the groups $G_4(VII)-G_4(VIII)$}.

If we introduce the functions:
$$st(u^\nu), \quad ct(u^\nu),\quad tn(u^\nu) =\frac{st(u^\nu)}{ct(u^\nu)}, \quad ctn(u^\nu)=tn^{-1}(u^\nu)$$
such that
$$
(st(u^\nu))_{,\nu}=ct(u^\nu),\quad(ct(u^\nu))_{,\nu}=-e_\nu st(u^\nu)\quad
(ct(u^\nu))^2+e_\nu (st(u^\nu))^2=(e_\nu)^2=1, \quad
$$
then expressions for metrics of spaces admitting these groups and operators of these groups can be combined.

Metrics:
$$
ds^2=a_1(e(du^1)^2ct(u^2)+(du^2)^2) + a_2(u^3)^2  +e_0(du^0)^2 \quad e=e_1e_2.
$$

Group operators:
$$
X_1 =p_1e_2tn(u^2)st(u^1)+ct(u^1)p^2,\quad   X_2 =-e_1(X_1)_{,1} ,\quad  X_3 =p_3, \quad X_4 = p_1.
$$
Structural constants:
$$
 C^A_{1}=-e\delta^A_4,\quad C^A_{14}=e_1\delta^A_2, \quad C^A_{24}=-\delta^A_1 \quad \rightarrow \quad
 \tilde{C}^\alpha_{12}=ctn(u^2)(-st(u^1)\delta^\alpha_1+ct(u^1)\delta^\alpha_2)
$$
Matrix $\hat{\lambda}$:
$$
||\lambda^\alpha_\beta||=
\begin{pmatrix}ctn(u^2)st(u^1)e  &-ctn(u^2)ct(u^1)e  & 0 \\
   ct(u^1) & st(u^1)e^1  & 0  \\0 & 0 &1
\end{pmatrix}.\quad
$$
Functions $\omega^\alpha$:
$$
\omega^\alpha=ctn(u^2)(e\delta^\alpha_1 st(u^1)-ct(u^1)\delta^\alpha_2)+\delta^\alpha_3.
$$
Matrix $\hat{W}$:
$$
||W^\gamma_\alpha||=
\begin{pmatrix}e\omega^1\omega^2  &e(\omega^2)^2  & 0 \\
 -1-e(\omega^1)^2  & -e\omega^1\omega^2 & 0  \\0 & 0 &0
\end{pmatrix}.\quad
$$
From the system of equations \eqref{40} it follows:
$$
\mathbf{A}_{1}=\mathbf{A}_{2} = 0.
$$
Taking this into account the \eqref{36} will take the form:
$$
\mathbf{A}_{3,\alpha}=0 \quad \rightarrow \quad \mathbf{A}_{3}=-a_0.
$$
The components of the potential of the admissible electromagnetic field are as follows:
$$
A_{0}= A_{1}=A_{2}=0, \quad A_{3}=a_{0}.
$$

\subsection{The groups \boldmath{$G_4$} act transitively on a null subspace $V^{*}_3$}

\subsubsection{Group \boldmath{$G_4(I)$}}.

Metric:
$$
ds^2=a_1 \exp{(-2u^3)}(2du^1du^0+(du^2)^2) + a_2(u^3)^2) + e_0(du^0)^2,
$$
Group operators:
$$
X_1 =p_1,\quad   X_2 = p_2 ,\quad  X_3 =2u^1p_1 +u^2p_2+p_3,
\quad X_4 = u^2p_1 -u^0p_2.
$$
(( Only for this case the condition \quad $(\xi^\alpha_A)_{,0}\ne 0$ \quad arises.))
Structural constants:
$$
 C^A_{13}=2\delta^A_1,\quad C^A_{23}=\delta^A_2, \quad C^A_{24}=\delta^A_1,\quad C^A_{43}=2\delta^A_4 \quad \rightarrow \quad  \tilde{C}^\alpha_{43}=2\omega^\alpha=2(u^2\delta^\alpha_1 -u^0\delta^\alpha_2).
$$
Matrix $\hat{\lambda}$:
$$
||\lambda^\alpha_\beta||=
\begin{pmatrix}1 & 0 & 0 \\
   0 & 1 & 0  \\-2u^1&-u^2 &1
\end{pmatrix}.\quad
$$
Functions $\omega^\alpha$:
$$
\omega^\alpha=u^2\delta^\alpha_1-u^3\delta^\alpha_2.
$$
Matrix $\hat{W}$:
$$
W^\gamma_i=-\delta^{\gamma}_2\delta_{i0} - \delta^{\gamma}_1\delta_{i2} + u^0\delta^{\gamma}_2\delta_{i3}.
$$
From the system of equations \quad $W^\gamma_i\mathbf{A}_{\gamma}=0$ \quad it follows:
$$
\mathbf{A}_{1}=\mathbf{A}_{2} = 0.
$$
Taking this into account the \eqref{36} will take the form:
$$
\mathbf{A}_{3,\alpha}=0 \quad \rightarrow \quad \mathbf{A}_{3}=-a_0.
$$
The components of the potential of the admissible electromagnetic field are as follows:
$$
A_{0}= A_{1}=A_{2}=0, \quad A_{3}=a_0.
$$

\subsubsection{Group \boldmath{$G_4$}}.

Metric:
$$
ds^2=4du^3du^0 +a\exp{(-2u^3)}((du^1)^2+(du^2)^2) ,
$$
Group operators:
$$
X_1 =p_1,\quad   X_2 = p_2 ,\quad  X_3 =u^1p_1 +u^2p_2+p_3,
\quad X_4 = u^1p_2 -u^2p_1.
$$
Structural constants:
$$
 C^A_{13}=-\delta^A_1,\quad C^A_{23}=\delta^A_2, \quad C^A_{14}=\delta^A_2, \quad C^A_{24}=-\delta^A_1,
$$
Matrix $\hat{\lambda}$:
$$
||\lambda^\alpha_\beta||=
\begin{pmatrix}1 & 0 & 0 \\
   0 & 1 & 0  \\-2u^1&-u^2 &1
\end{pmatrix}.\quad
$$
Functions $\omega^\alpha$:
$$
\omega^\alpha=u^1\delta^\alpha_2-u^2\delta^\alpha_1.
$$
Matrix $\hat{W}$:
$$
W^\gamma_\alpha=-\delta^{\gamma}_2\delta_{\alpha 1} + \delta^{\gamma}_1\delta_{\alpha2} + (u^2\delta^{\gamma}_1+u^1\delta^{\gamma}_2)\delta_{\alpha3}.
$$
From the system of equations \quad $W^\gamma_i\mathbf{A}_{\gamma}=0$ \quad it follows:
$$
\mathbf{A}_{1}=\mathbf{A}_{2} = 0.
$$
Taking this into account the \eqref{36} will take the form:
$$
\mathbf{A}_{3,\alpha}=0 \quad \rightarrow \quad \mathbf{A}_{3}=-a_0.
$$
The components of the potential of the admissible electromagnetic field are as follows:
$$
A_{0}= A_{1}=A_{2}=0, \quad A_{3}=a_0.
$$

\subsubsection{Group $G_4(V)$}

Metric:
$$
ds^2=2du^3du^0 +a(\exp{(-2u^3)}(du^2)^2+(du^2)^2)+\varepsilon\exp{(-u^3)}du^0 du^2 \quad \varepsilon = 0, 1.
$$
Group operators:
$$
X_1 =p_1,\quad   X_2 = p_2 ,\quad  X_3 =u^2p_2+p_3,
\quad X_4 =-\varepsilon p_1\exp{-u^3} +((u^2)^2+\exp{-2u^3})p_2 + 2u^2p_3.
$$
Structural constants:
$$
C^A_{23}=\delta^A_2, \quad C^A_{24}=2\delta^A_3, \quad C^A_{34}=2u^2\delta^A_3 -\delta^A_2((u^2)^2+\exp{2u^2}),
$$
Matrix $\hat{\lambda}$:
$$
||\lambda^\alpha_\beta||=
\begin{pmatrix}1 & 0 & 0 \\
   0 & 1 & 0  \\0 &-u^2 &1
\end{pmatrix}.\quad
$$
Functions $\omega^\alpha$:
$$
\omega^\alpha=-\delta^\alpha_1 \exp{u^3}-((u^2)^2+\exp{-2u^3})\delta^\alpha_2+2u^2\delta^\alpha_3.
$$
Matrix $\hat{W}$:
$$
W^\gamma_\beta=-2(\delta^{\gamma}_3 + u^2\delta^{\gamma}_2)\delta_{\beta 2} - (2u^2\delta^{\gamma}_3 -\varepsilon\delta^{\gamma}_1 \exp{u^3})\delta_{\beta3}.
$$
From the system of equations \quad $W^\gamma_i\mathbf{A}_{\gamma}=0$ \quad it follows:

\begin{equation}\label{44}
\mathbf{A}_{3}+u^2\mathbf{A}_{2} = 0, \quad 2u^2\mathbf{A}_{3} -\varepsilon\mathbf{A}_{1}=0.
\end{equation}

Consider separately the variants \quad $\varepsilon=0$ \quad and \quad $\varepsilon=1$.

A) \quad $\varepsilon=0$.\quad  In this case, the system \eqref{36} will take the form:
$$
\mathbf{A}_{3,\alpha}=0 \quad \rightarrow \quad \mathbf{A}_{3}=-a_0.
$$
The components of the potential of the admissible electromagnetic field are as follows:
$$
A_{0}= A_{1}=A_{2}=0, \quad A_{3}=a_0.
$$

B) \quad $\varepsilon=1$. \quad  From \eqref{44} it follows:
\begin{equation}\label{45}
\mathbf{A}_{2}=-\frac{\mathbf{A}_{1}\exp{u^3}}{2(u^2)^2}, \quad
\mathbf{A}_{3} = \frac{\mathbf{A}_{1}}{u^2}.
\end{equation}
From the system \eqref{36} it follows:
$$
\mathbf{A}_{1}=a_0,\quad \mathbf{A}_{2,2},
$$
which contradicts the relations \eqref{45}. Thus, in this case, there is no permissible electromagnetic field.

\subsubsection{Group $G_4(VIII)$}.

Metric:
$$
ds^2=a_1((du^1)^2+\sin^2{u^1}(du^2)^2) + 2a_2\varepsilon\cos{u^1}du^1du^0 +e_0(du^0)^2 \quad \varepsilon = 0, 1.
$$

Group operators:
$$
X_1 =p_1,\quad   X_2 = p_2 ,\quad  X_3 =-\frac{\cos{u^3} \sin{u^1}}{\sin{u^3}}p^1 + \frac{\varepsilon\sin{u^1}}{\sin{u^3}}p_2 + \cos{u^1} p_3,\quad X_4 = \frac{\partial{X_3}}{\partial{u^2}}.
$$

Structural constants:
$$
 C^A_{13}=\delta^A_4  \rightarrow
  \tilde{C}^\gamma_{13}=\omega^\gamma, \quad C^A_{41}=\delta^A_3, \quad C^A_{34}=\delta^A_1.
$$

Matrix $\hat{\lambda}$:
$$
||\lambda^\alpha_\beta||=
\begin{pmatrix}1 & 0 & 0 \\
   0 & 1 & 0  \\\frac{\sin{u^1}\cos{u^3}}{\cos{u^1}\sin{u^3}}& -\frac{\varepsilon\sin{u^1}}{\cos{u^1}\sin{u^3}} &\frac{1}{\cos{u^1}}
\end{pmatrix},\quad
$$

Functions $\omega^\alpha$:
$$
\omega^\alpha=-\frac{\delta^\alpha_1\cos{u^3}}{\cos{u^1}\sin{u^3}}
+\frac{\varepsilon\delta^\alpha_2}{\cos{u^1}\sin{u^3}} - \frac{\delta^\alpha_3 \sin{u^1}}{\cos{u^1}}.
$$

Elements of the matrix $W^\alpha_\beta$:
$$
W^\alpha_\gamma=\delta_{\alpha1}(\delta^\gamma_3 + \omega^3\omega^\gamma)-
\delta_{\alpha3}(\delta^\gamma_1 + \omega^1\omega^\gamma).
$$
These formulas differ from the formulas given in the variant \quad 6  \quad (when the group \quad $G_4(VIII)$\quad acts on a non-isotropic hypersurface \quad $V_3$\quad), \quad by the presence of the $\varepsilon$ quantity. Repeating the calculations performed in variant \quad $6$,\quad
we obtain the solution of the system \quad $W^\gamma_\alpha \mathbf{A}_\gamma=0 $ \quad in the following form:

\begin{equation}\label{47}
\mathbf{A}=-\varepsilon\mathbf{A}_{2}\sin{u^3}\cos{u^1} \quad
\mathbf{A}_{1}=\varepsilon\mathbf{A}_2\cos{u^3},\quad \mathbf{A}_{3}=\varepsilon\mathbf{A}_2\sin{u^1}\sin{u^3}.
\end{equation}

If \quad $\varepsilon=1,$ \quad From the system \eqref{41} we get
$$
\mathbf{A}_{2,\alpha} = 0 \quad \rightarrow \quad \mathbf{A}_{2} = a_0,
$$
And the components of the potential of the admissible electromagnetic field are as follows:
$$
A_{0}= A_{3}=0, \quad  A_{2}=a_0, \quad A_{1}=a_{0}\cos{u^3}.
$$
If \quad $\varepsilon=0,$ \quad from \eqref{41} and \eqref{47} it follows:
$$ A_i=0.$$
\section{Conclusion}
The classification of admissible electromagnetic fields carried out in this article should be considered as a stage in the general program of research into the problem of integrating the classical and quantum equations of motion of a test particle in external fields of different nature in spaces with symmetry due to the sets of Killing fields. This program is effectively implemented for the Stackel spaces. Complete or partial classifications of the Stackel and special Stackel spaces of the electrovacuum are obtained in the General Theory of Relativity, as well as in the scalar-tensor theory.

The possibility of applying the theory of symmetry to the construction of cosmological models in the theory of gravity, including the Brans---Dicke theory, is also being studied (see, for example, \cite{Odintsov} -- \cite{A}). As is known, from a physical point of view, the Robertson---Walker space is the most important special case of the Stackel space.

    We also note the activity on the study of spaces belonging to the intersection of Stackel sets and homogeneous spaces (see \cite{Osetrin1}-\cite{Osetrin4}). The solution of the classification problem considered in the program allows us to extend this activity to solving problems in the presence of admissible electromagnetic fields.

As a rule, in the given examples, the classification problem was considered within the framework of a specific theory of gravity with the involvement of the gravitational field equations. This greatly simplifies the classification since it imposes additional serious restrictions on the potential and metric. In \cite{VVO2}-\cite{VVO4} the classification of Stackel spaces and admissible external electromagnetic fields, in which the Hamilton---Jacobi equation for a charged test particle allows complete separation of variables, without involving the field equations, was first carried out. At the same time, the same classification problem for the Klein---Gordon---Fock equation is far from being solved.

 A fundamentally different situation takes place in the implementation of the classification of admissible electromagnetic fields in spaces admitting groups of motions, since in these spaces the algebras of integrals of motion are already known, due to the theorem proved in this paper. To solve the problem, it remains to integrate the systems of equations \eqref{13} for each space with a Lorentzian signature that admits a group of motions. This classification problem is expected to be completely solved in the near future.

 At the next stage of research, within the framework of the general program, one can consider the squared Dirac---Fock equations. In \cite{Bagrov},\cite{Mac} a method for integrating the Dirac---Fock equation was proposed, which allows one to reduce the solution of some bispinor equations to the problem of integrating linear scalar equations of the second order. It is supposed to investigate these scalar equations for the existence of admissible electromagnetic fields.

\quad

\bf{Acknowledgments}
The work is partially supported by the Ministry of Education of the Russian Federation, Project No. FEWF-2020-003.

\quad


\end{document}